\documentclass{article}

\usepackage[preprint]{neurips_2019}

\usepackage[utf8]{inputenc} % allow utf-8 input
\usepackage[T1]{fontenc}    % use 8-bit T1 fonts
\usepackage{hyperref}       % hyperlinks
\usepackage{url}            % simple URL typesetting
\usepackage{booktabs}       % professional-quality tables
\usepackage{amsfonts}       % blackboard math symbols
\usepackage{nicefrac}       % compact symbols for 1/2, etc.
\usepackage{microtype}      % microtypography

\usepackage{amssymb}
\usepackage{amsmath}
\usepackage{bm}
\usepackage{algorithm}
\usepackage{algpseudocode}
\usepackage{comment}
\usepackage{graphicx}
\usepackage{natbib}

\title{A Complementary Learning Systems Approach to Temporal Difference Learning}

\author{%
	Sam Blakeman\\
	Centre for Brain and Cognitive Development \\
	Department of Psychological Sciences\\
	Birkbeck, University of London \\
	Malet Street, WC1E 7HX UK \\
	\texttt{sblake03@mail.bbk.ac.uk} \\
	\And
	Denis Mareschal \\
	Centre for Brain and Cognitive Development \\
	Department of Psychological Sciences\\
	Birkbeck, University of London \\
	Malet Street, WC1E 7HX UK \\
	\texttt{d.mareschal@bbk.ac.uk} \\
}

\begin{document}
	
\maketitle

\begin{abstract}
Complementary Learning Systems (CLS) theory suggests that the brain uses a 'neocortical' and a 'hippocampal' learning system to achieve complex behavior. These two systems are complementary in that the 'neocortical' system relies on slow learning of distributed representations while the 'hippocampal' system relies on fast learning of pattern-separated representations. Both of these systems project to the striatum, which is a key neural structure in the brain's implementation of Reinforcement Learning (RL). Current deep RL approaches share similarities with a 'neocortical' system because they slowly learn distributed representations through backpropagation in Deep Neural Networks (DNNs).  An ongoing criticism of such approaches is that they are data inefficient and lack flexibility. CLS theory suggests that the addition of a 'hippocampal' system could address these criticisms. In the present study we propose a novel algorithm known as Complementary Temporal Difference Learning (CTDL), which combines a DNN with a Self-Organising Map (SOM) to obtain the benefits of both a 'neocortical' and a 'hippocampal' system. Key features of CTDL include the use of Temporal Difference (TD) error to update a SOM and the combination of a SOM and DNN to calculate action values. We evaluate CTDL on grid worlds and the Cart-Pole environment, and show several benefits over the classic Deep Q-Network (DQN) approach. These results demonstrate (1) the utility of complementary learning systems for the evaluation of actions, (2) that the TD error signal is a useful form of communication between the two systems and (3) the biological plausibility of the proposed approach.
\end{abstract}

\section{Introduction}

Reinforcement Learning (RL) \citep{Sutton1998} represents a computational framework for modelling complex reward-driven behavior in both artificial and biological agents. For cognitive scientists it is of continuing interest to explore how RL theory maps onto neural structures in the brain \citep{Niv2009, Lee2012}. One of the most influential findings in this regard is the encoding of Temporal Difference (TD) error by phasic midbrain dopaminergic neurons \citep{Schultz1997, Schultz2016}. One of the major projection sites of these neurons is the striatum \citep{Doherty2003, Mcclure2003, Bray2007} and it has been proposed that the striatum is responsible for evaluating states and actions for decision making \citep{Schultz1992, Houk1995, Schultz1998, Setlow2003, Roesch2009}. Interestingly, the striatum receives inputs from both cortical areas and the hippocampus, suggesting that it is responsible for evaluating different forms of information.

Complementary Learning Systems (CLS) theory posits that the neocortex and hippocampus have complementary properties that allow for complex behavior \citep{McClelland1995, Kumaran2016}. More specifically, the hippocampus relies on fast learning of conjunctive, pattern-separated memories. These memories then support the learning of a second system, the neocortex, which slowly learns distributed representations that support generalization across features and experiences. The purpose of the present study is to explore how the brain's RL machinery might utilise these opposing properties to achieve complex behavior. 

Much of the recent success of RL has been due to the combination of classical RL approaches with the function approximation properties of Deep Neural Networks (DNNs), known as deep RL \citep{Francois-lavet2018}. Typically in deep RL, the action-value function $Q(s, a)$ is represented using a DNN that takes the state $s_t$ as input and outputs the corresponding action values for that state. Despite impressive results, such as super human-level performance on Atari video games \citep{Mnih2015}, deep RL approaches are often criticised for being data inefficient and adapting poorly to changes in the input distribution \citep{Lake2017}. From a CLS perspective, the DNNs used in deep RL can be seen as sharing many commonalities with a 'neocortical' learning system. In particular, both the neocortex and DNNs rely on small learning rates and distributed representations for efficient generalization. 

CLS theory suggests that the addition of a 'hippocampal' learning system to deep RL approaches may improve our understanding of how RL is implemented in the brain and address the aforementioned criticisms of deep RL. Ideally an agent should be able to utilize the advantages of a 'neocortical' system (distributed representations and generalization) and a 'hippocampal' system (fast learning and pattern-separation) to perform complex reward-driven behavior. Indeed, many theoretical advantages have been proposed for the use of hippocampal episodic information in RL. In particular, it has been suggested that episodic information can be used to approximate value functions, increase data efficiency and reconcile long-range dependencies \citep{Gershman2017}.

An alternative to using a DNN to represent the action-value function is to represent it in a tabular manner \citep{Sutton1998}. Such an approach is more in line with a hippocampal learning system as experiences are stored in a pattern-separated manner and larger learning rates can be used. Importantly, the tabular case means that every action value is stored as its own memory, which eliminates the potential for interference. This is in contrast to DNNs that naturally suffer from interference due to their distributed representations. However, as the number of states and/or actions increases, the tabular case will require more experience to encounter each action-value and more computational resources to store the values. The distributed representations of DNNs then become advantageous because they allow for efficient generalization over the state space. In an ideal scenario a DNN would be responsible for generalization over certain areas of the state space while a tabular method would store pattern-separated memories that are crucial to behavior and that violate the generalizations of the network.

Previous work in deep RL has often touched upon CLS theory and the benefits of a hippocampal learning system. Indeed, one of the most influential deep RL approaches, the Deep Q-Network (DQN) \citep{Mnih2015}, utilises a secondary system that tentatively mirrors a hippocampal learning system. More specifically, the DQN has a table that stores past experiences in a pattern-separated manner and then uses them to train a DNN in an interleaved fashion. This process is tentatively compared to ‘replay’; a biological phenomenon that appears to replay information from the hippocampus to the neocortex in biological agents \citep{Olafsdottir2018}. However, despite the DQN having what appears to be two complementary learning systems, the decision making (calculation of Q values) is ultimately based on the predictions of the DNN, which learns slowly via distributed representations.

More recently, research in deep RL has begun to demonstrate some of the advantages of an explicit 'hippocampal' learning system that evaluates states and actions \citep{Botvinick2019}. Most notably \citet{Blundell2016} proposed an algorithm called 'model-free episodic control', which consisted of a table containing the maximum return (sum of discounted rewards) for each state-action pair experienced. The memory requirements for this table were kept constant by removing the least recently updated table entry once the size limit had been reached. Each observation from the environment was projected by an embedding function (either a random projection or a variational autoencoder) to a state value and actions were selected based on a k-nearest neighbours method, which allowed for some degree of generalization to novel states. \citet{Blundell2016} tested this approach on the Arcade Learning Environment (Atari) \citep{Bellemare2013} and Labyrinth \citep{Mnih2016}, which both require the use of visual information to learn an optimal policy. The results of these simulations showed that model-free episodic control was significantly more data efficient than other classical deep RL approaches, suggesting that episodic information is important for fast learning. 

While taking a first step towards highlighting the benefits of a 'hippocampal' learning system that utilises fast learning of pattern-separated information, the work of \citet{Blundell2016} has several notable drawbacks. Firstly, the table recorded the maximum return from any given episode and used this to inform the policy of the agent. This naturally cannot handle stochastic environments, where the expected return is the important quantity and not the maximum return of an individual episode. Secondly, this approach is likely to be highly inflexible. For example if a state-action pair suddenly becomes highly aversive then the entry in the table will not be updated because only the maximum value is stored. A third criticism is that the approach relies on the full return for each state-action pair and this is only possible when the task has distinct finite episodes. Some of these criticisms have been addressed in subsequent work, for example \citet{Pritzel2017} propose a fully differentiable version of 'model-free episodic control' that learns the embedding function in an online fashion using N-step Q-learning.

Not withstanding, what is most pertinent to the present study is that 'model-free episodic control', and its various derivatives, do not rely on two complementary learning systems that operate in parallel to evaluate actions. The embedding function may be tentatively compared to a 'neocortical' learning system but it operates before the 'hippocampal' learning system and as a result only the output of the 'hippocampal' learning system is used to evaluate action values. This means that any advantages that may be conferred from the additional predictions of a 'neocortical' learning system are lost. In essence, the aforementioned approaches cannot arbitrate between the predictions of a 'neocortical' and a 'hippocampal' learning system, but are instead restricted to using episodic predictions. This is inconsistent with the finding that the striatum receives inputs from both cortical areas and the hippocampus and needs to arbitrate between the two \citep{Pennartz2011}.

With these criticisms in mind, we present a novel method for imbuing a deep RL agent with both a 'neocortical' and a 'hippocampal' learning system so that it benefits from both types of learning system. Most importantly these two systems: (1) learn in parallel, (2) communicate with each other using a biologically plausible signal, and (3) both make action value predictions. We represent the 'neocortical' system as a DNN and the 'hippocampal' system as a Self-Organizing Map (SOM). Importantly, the size of the SOM is significantly smaller than the state space experienced by the agent so as to replicate the restricted computational resources experienced by biological agents. The SOM is tasked with storing pattern-separated memories of states that the DNN is poor at evaluating. To achieve this we use the TD error from a DNN in order to train the SOM. Critically, this novel CLS approach demonstrates how the TD error of a 'cortical' system can be used to inform a 'hippocampal' system about when and what memories should be stored, with both systems contributing to the evaluation of action-values. This allows the agent to utilize the benefits of both a neocortical and hippocampal learning system for action selection. We call our novel algorithm Complementary Temporal Difference Learning (CTDL) and demonstrate that it can improve the performance and robustness of a deep RL agent in a simple grid world and on the Cart-Pole task.

\section{Methods}

\subsection{The Reinforcement Learning Problem}

The general goal of a reinforcement learning agent is to select actions based on perceived states in order to maximise future expected rewards. In simple terms the agent chooses an action $a_t$ given a state $s_t$. The environment then responds to this decision and produces the next state $s_{t+1}$ and a reward $r_{t+1}$. This agent-environment loop continues either indefinitely or until a terminal state is reached. Typically future rewards are discounted so that more immediate rewards are worth more than distant rewards. This is done using a discount factor $\gamma \in (0, 1)$ which is applied at each time step. The return $R_t$ at time $t$ is defined as the discounted sum of future rewards:
\begin{equation}
R_t = \sum_{t' = t}^{T} \gamma^{t'-t}r_{t'}
\end{equation}
Where $r_{t'}$ is the reward value at time $t'$ and $T$ is the time-step at which the task or episode finishes. Solving the RL problem equates to learning a policy $\pi$ that maps from states to actions ($\pi : s \mapsto a$) and achieves the greatest possible expected return from every state. This is known as the optimal policy $\pi^*$. One possible method of finding $\pi^*$ is to learn the optimal action-value function $Q^*(s, a)$, which provides the expected return of taking action $a$ in state $s$ and following $\pi^*$ thereafter. 
\begin{equation}
\begin{split}
Q^*(s,a) &= \max_\pi Q^\pi(s,a) \\
&= \max_\pi \mathbb{E}_\pi[R_t \mid s_t = s, a_t = a] \\
\end{split}
\end{equation}
Once an agent has learnt the optimal action-value function it can act optimally by picking the action with the largest Q value given the state it is in ($\arg\max_a Q(s,a)$). Importantly the optimal action-value function can be defined recursively using the Bellman equation:
\begin{equation}
Q^*(s,a) = \mathbb{E}_{s'}[r + \gamma \max_{a'}Q^*(s', a') | s, a]
\end{equation}
This recursive definition of the action-value function forms the basis for many learning algorithms. One such algorithm is Q-learning, which is a form of Temporal Difference (TD) learning. Q-learning utilises the following update rule to learn the optimal action-value function:
\begin{equation}
Q(s_t,a_t) \gets Q(s_t,a_t) + \alpha[r_{t+1} + \gamma \max_a Q(s_{t+1}, a) - Q(s_t,a_t)]
\end{equation}
Where $[r_{t+1} + \gamma \max_a Q(s_{t+1}, a) - Q(s_t,a_t)]$ is known as the TD error, which has been proposed to exist in biological agents \citep{Schultz1997, Doherty2003}. Importantly the action-value function can be represented in a tabular manner or with a function approximator such as a DNN. In the case of a DNN, the state is provided as input and each output unit corresponds to the value of a single action. The parameters of the network $\theta$ are updated so that $Q(s,a)$ moves closer to $r + \gamma \max_a Q(s_{t+1}, a)$. In order to achieve this the objective function of the neural network is set to the mean squared error between the two values i.e. the mean square of the TD error:
\begin{equation}
J(\theta) = \mathbb{E}_{s_t, a_t, r_{t+1}, s_{t+1}}[((r_{t+1} + \gamma \max_{a}Q(s_{t+1}, a; \theta)) - Q(s_t, a_t; \theta))^2]
\end{equation}

\subsection{Complementary Temporal Difference Learning (CTDL)}

Our novel approach combines a DNN with a SOM to imbue an agent with the benefits of both a 'neocortical' and 'hippocampal' learning system. The DNN is a simple feed-forward network that takes the current state as input and outputs the predicted action values for each action. The network is trained using the same training objective as \citet{Mnih2015} and a copy of the network is made every $C$ time steps in order to improve training stability. The optimizer used was RMSProp and the hyper-parameter values can be seen in Table \ref{table:HyperParameterTable}. Importantly, unlike in \citet{Mnih2015}, no memory buffer is used to record past experiences, which saves considerable memory resources. The SOM component is represented as a square grid of units, with each unit having a corresponding action-value $Q(u, a)$ and weights $\beta_u$ that represent a particular state.

A general outline of the algorithm detailing how the DNN and SOM interact can be seen in Algorithm \ref{alg:CTDLAlgorithm}. In simple terms, the TD error produced by the DNN is used to update the SOM and both systems are used to calculate Q values for action selection. When the agent observes the state $s_t$, the closest matching unit in the SOM $u_t$ is calculated based on the euclidean distance between the units weights $\beta_u$ and $s_t$. This distance is also used to calculate a weighting parameter $\eta \in \{0, 1\}$, which is used to calculate a weighted average of the action values from the SOM and the DNN. If the best matching unit is close to the current state then a larger weighting will be applied to the Q value produced by the SOM. A free parameter $\tau_{\eta}$ acts as a temperature parameter to scale the euclidean distance between $\beta_u$ and $s_t$ when calculating the weighted average.

For learning in both the DNN and the SOM, the TD error is calculated using the difference between the target value and the predicted Q value of the DNN. The TD error is used to perform a gradient descent step with respect to the parameters $\theta$ of the DNN, which ensures that the predictions of the DNN move towards the weighted average of the SOM and DNN predictions. After updating the DNN, the TD error is also used to update the SOM. More specifically, the TD error is used to create an exponentially increasing value $\delta \in \{0, 1\}$, which scales the standard deviation of the SOM's neighbourhood function and the learning rate of the SOM's weight update rule. Again a temperature parameter $\tau_{\delta}$ is used to scale the TD error. Next, the action value of the closest matching unit from the previous time step $Q^{SOM}(u_{t-1}, a_{t-1})$ is updated using the learning rate $\rho$, the weighting from the previous time step $\eta_{t-1}$ and the difference between $Q^{SOM}(u_{t-1}, a_{t-1})$ and the target value $y_t$. The inclusion of $\eta_{t-1}$ ensures that the action value only receives a large update if the closest matching unit is similar to the state value.

To aid in the training of the DNN and to mimic biological 'replay', the contents of the SOM are replayed to the DNN as a training batch for gradient descent. To construct the training batch the actions $a_t$ are sampled randomly, the states $s_t$ are set to a random sample of the SOM weights $\beta_u$ and the target values $y_t$ are set to the corresponding Q values stored in the SOM. Finally, the agents actual action is chosen in an $\epsilon$-greedy manner with respect to the weighted average of the predicted DQN and SOM Q values.

The aforementioned algorithm has several interesting properties. Firstly, the calculation of Q values involves the contribution of both the DNN and the SOM. The size of their respective contributions are controlled by the parameter $\eta$, which ensures that if the current state is close to one stored in SOM memory then the Q value predicted by the SOM will have a larger contribution. This is akin to retrieving a closely matching episodic memory and using its associated value for action selection. Secondly, because the SOM is updated using the TD error produced by the DNN, it is biased towards storing memories of states that the DNN is poor at evaluating. Theoretically this should allow the DNN to learn generalizations across states, while the SOM picks up on violations or exceptions to these generalizations and stores them in memory along with a record of their action values. If after many learning iterations the DNN converges to a good approximation of the optimal action-value function then no TD error will be produced and the SOM will be free to use its resources for other tasks. Finally, the SOM can use much larger learning rates than the DNN because it relies on a tabular approximation of the action-value function, which should improve data efficiency.
\begin{algorithm}
	\caption{Complementary Temporal Difference Learning (CTDL)}
	\begin{algorithmic}
		\State{Initialize probability of selecting a random action $\epsilon$ = 1}
		\State{Initialize SOM weights $\beta$ to random locations in the grid world}
		\State{Initialize SOM action-values $Q^{SOM} = 0$}
		\State{Initialize action-value function $Q^{DNN}$ with random weights $\theta$}
		\State{Initialize target action-value function $\tilde{Q}^{DNN}$ with weights $\theta^- = \theta$}
		\For{$e= 1, E$}
		\State{If $\epsilon > \epsilon^{end}$ then decrease $\epsilon$ by $\epsilon^{end} / E^{\epsilon}$}
		\For{$t=1, T$}
		\State{Observe current state $s_t$ and reward $r_t$}
		\State{Retrieve SOM unit $u_t$ that is closest to $s_t$}
		\State{$u_t =\arg\min_u ||\beta_u - s_t||^2$}
		\State{Calculate weighting $\eta$ based on distance}
		\State{$\eta = \exp{(-||\beta_{u_t} - s_t||^2 / \tau_{\eta})}$ }
		\State{Calculate $Q(s_t, a')$ as weighted average of SOM and DNN values}
		\State{$Q(s_t, a') = \eta Q^{SOM}(u_t, a') + (1 - \eta)\tilde{Q}^{DNN}(s_t, a'; \theta^-)$}
		\State{
			\[ 
			\text{set } y_t = 
			\begin{cases} 
			r_t & \text{if episode is over} \\
			r_t + \gamma \max_{a'}{Q(s_t, a')} & \text{otherwise}
			\end{cases}
			\]	
		}
		\State{Perform gradient descent step on $(y_t - Q^{DNN}(s_{t-1}, a_{t-1}; \theta))^2$ with} 
		\State{respect to the network parameters $\theta$}
		\State{Calculate $\delta$ based on the TD error produced by the DNN}
		\State{$\delta = \exp{(|y_t - Q^{DNN}(s_{t-1}, a_{t-1}; \theta)| / \tau_{\delta})} - 1$}
		\State{Calculate the neighbourhood function based on $u_{t-1}$}
		\State{$T_{u_j, u_{t-1}} = \exp{(- ||l_{u_j} - l_{u_{t-1}}||^2 / 2(\sigma_c + (\delta * \sigma)))}$}
		\State{Update the weights $\beta$ of SOM}
		\State{$\Delta \beta_{ji} = \alpha * \delta * T_{u_j, u_{t-1}} (s_{t-1, i} - \beta_{ji})$}
		\State{Update the action value $\Delta Q^{SOM}(u_{t-1}, a_{t-1}) =$}
		\State{$\rho * \eta_{t-1} * (y_t - Q^{SOM}(u_{t-1}, a_{t-1}))$}
		\State{Replay contents of SOM to DNN using a random sample of actions}
		\State{$a_t$ and unit weights $\beta_u$. $y_t$ is set to $Q^{SOM}(\beta_u, a_t)$}
		\State{Select random action with probability $\epsilon$, else $a_t = $
			\State{$\arg\max_{a'}\eta Q^{SOM}(u_t, a') + (1 - \eta)Q^{DNN}(s_t, a'; \theta)$}}
		\State{Every $C$ steps reset $\tilde{Q}^{DNN} = Q^{DNN}$}
		\State{If the goal has been reached then \textbf{break} and end episode}
		\EndFor
		\EndFor
	\end{algorithmic}
	\label{alg:CTDLAlgorithm}
\end{algorithm}

\subsection{Simulated Environments}

\subsubsection{Grid World Task}

To evaluate our approach we generated a set of symmetric 2D grid worlds (Figure \ref{fig:F1_CTDLAndDQN}). Each cell in the grid world represents a state $s \in \mathbf{R}^2$ that is described by its $x$ and $y$ position. If $N$ is the number of cells in the grid world, then $\frac{N}{5}$ negative rewards (-1) are randomly placed in the grid world along with a single positive reward (+1) and the agents starting position. The agent's task is to reach the positive reward, at which point the episode is over and a new episode begins. The agent's action space is defined by four possible actions (up, down, left and right), each of which moves the agent one cell in the corresponding direction with probability 1. If the agent chooses an action that would move it out of the grid world then it remains where it is for that time step.

\subsubsection{Cart-Pole}

In addition to the grid world task we also evaluated CTDL on the Cart-Pole environment as provided by the OpenAI Gym \citep{Brockman2016}. The Cart-Pole problem consists of a cart with a pole attached by a single un-actuated joint. The goal of the agent is to control the velocity of the cart on a linear friction-less track so that the pole stays up-right. The state observed by the agent is made up of four values which correspond to the position of the cart $[-4.8, 4.8]$, the velocity of the cart $[-\infty, \infty]$, the angle of the pole $[\sim-41.8, \sim41.8]$ and the velocity of the end of the pole $[-\infty, \infty]$. Two actions are available to the agent; push the cart left and push the cart right. The agent receives a reward of +1 at every time step and an episode ends either when the angle of the pole is greater than 15 degrees, the cart moves off the screen or the episode length is greater than 500.

\section{Results}

In our first simulation we compare CTDL to the standard DQN described by \citet{Mnih2015} on a range of grid worlds. There are two key differences to highlight between these two approaches. Firstly, a standard DQN stores a memory buffer of size $N$ that is used to replay past experiences whereas CTDL relies on the contents of a SOM for replay. For our simulations we set the memory buffer size $M$ of the DQN to 100,000 while the size of the SOM was set to 36 units. This represents a significant decrease in memory resources between the two approaches. The second key difference is that a standard DQN only uses the DNN for calculation of Q values whereas CTDL also incorporates the predictions of a SOM. This allows CTDL to utilize the benefits of a 'hippocampal' learning system during decision making, namely pattern-separated memories and larger learning rates. Importantly, the DNN used in both CTDL and the DQN had the same architecture and training parameters. Both models learnt from 1000 episodes, with a maximum episode length of 1000. The probability of randomly selecting an action $\epsilon$ was linearly decreased from 1.0 to 0.1 over the first 200 episodes. Table \ref{table:HyperParameterTable} contains the values of all the hyper-parameters used in the simulations.

\begin{table}
	\caption{Hyper-parameter values used for all simulations. $\tau_{\eta}$, $\tau_{\delta}$, $\sigma$, $\sigma_c$, $\alpha$ and $\rho$ were selected by using a random grid search on a single grid world.}
	\label{table:HyperParameterTable}
	\centering
	\begin{tabular}{lll}
		\toprule
		Parameter & Value & Description \\
		\midrule
		$N$ & 100 & Size of grid world\\
		$E$ & 1,000 & Number of episodes for learning \\
		$\epsilon^{end}$ & .1 &  Final probability of selecting a random action\\ 
		$E^{\epsilon}$ & 200 &  Number of episodes to linearly decrease $\epsilon$ over\\
		$T$ & 1,000 & Maximum number of time steps per episode\\
		$\tau_{\eta}$ & 10 & Temperature for calculating $\eta$\\
		$\tau_{\delta}$ & 1 &  Temperature for calculating $\delta$\\
		$\sigma$ & .1 &  Standard deviation of the SOM neighbourhood function\\
		$\sigma_c$ & .1 &  Constant for denominator in SOM neighbourhood function\\
		$\alpha$ & .01 & Learning rate for updating the weights of the SOM\\
		$\rho$ & .9 & Learning rate for updating the Q values of the SOM\\
		$C$ & 10,000 & Number of steps before updating the target network\\
		$\lambda$ & .00025 & Learning rate for RMSProp\\
		$\kappa$ & .95 & Momentum for RMSProp\\
		$\phi$ & .01 & Constant for denominator in RMSProp\\
		$\gamma$ & .99 & Discount factor for future rewards\\
		\bottomrule
	\end{tabular}
\end{table} 

\begin{figure}
	\centering
	\includegraphics[height=14cm, width=11cm]{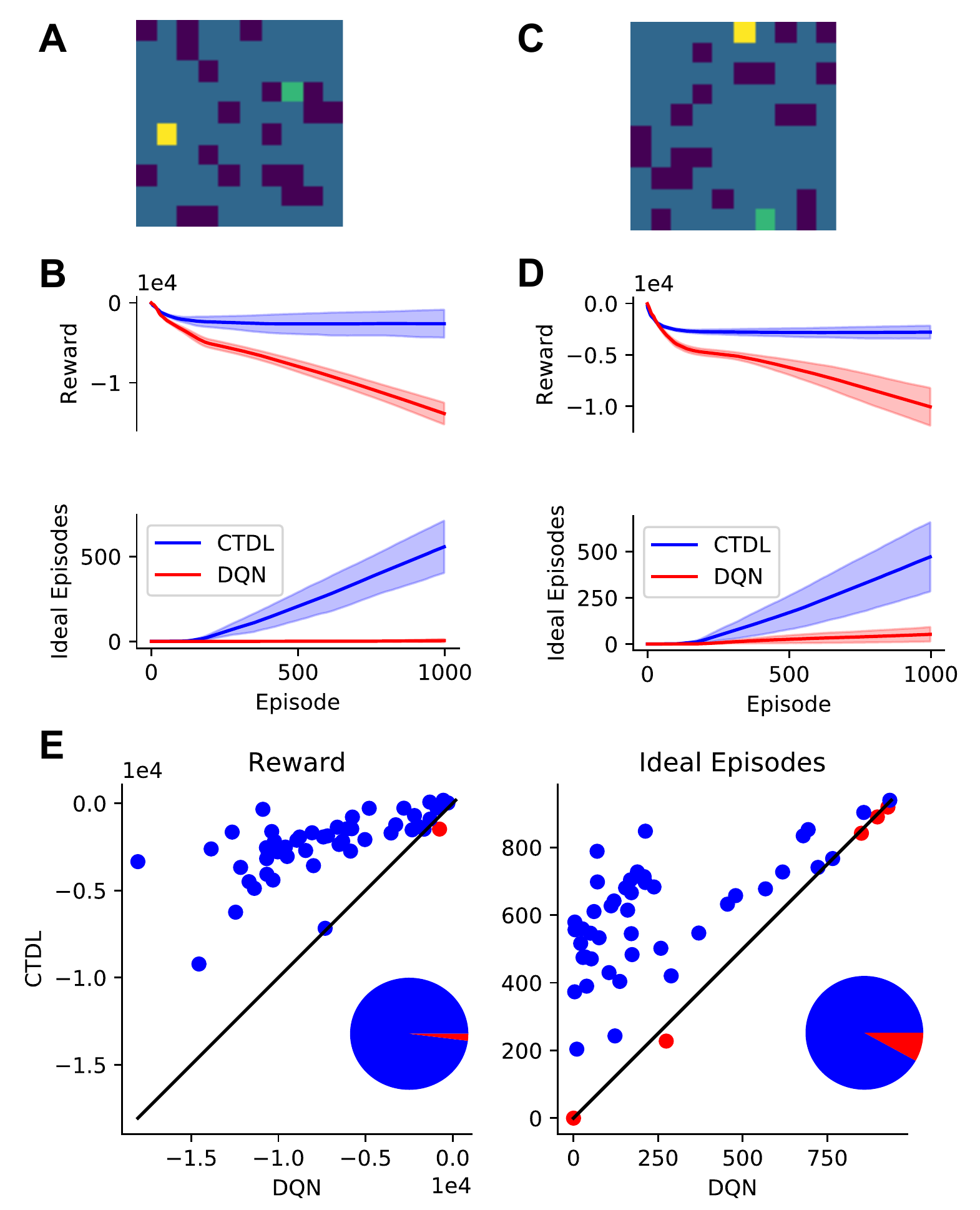}
	\caption{\textbf{A:} First example grid world, dark blue cells represent negative rewards (-1), the green cell represents the goal (+1) and the yellow cell represents the agents starting position. \textbf{B:} Performance of CTDL and DQN on the first example gird world in terms of cumulative reward and 'ideal' episodes over the course of learning. An 'ideal' episode is an episode where the agent reached the goal location without encountering a negative reward. Both CTDL and DQN were run 30 times on each maze. \textbf{C:} Second example gird world. \textbf{D:} Performance of CTDL and DQN on the second example grid world. \textbf{E:} Scatter plots comparing the performance of CTDL and DQN on 50 different randomly generated grid worlds. Both CTDL and DQN were run 30 times on each maze and the mean value at the end of learning was calculated. Blue points indicate grid worlds where CTDL out-performed DQN and red points indicate grid worlds where DQN out-performed CTDL. The pie charts to the lower right indicate the proportions of blue and red points.}
	\label{fig:F1_CTDLAndDQN}
\end{figure}

\begin{figure}
	\centering
	\includegraphics[height=7cm, width=12cm]{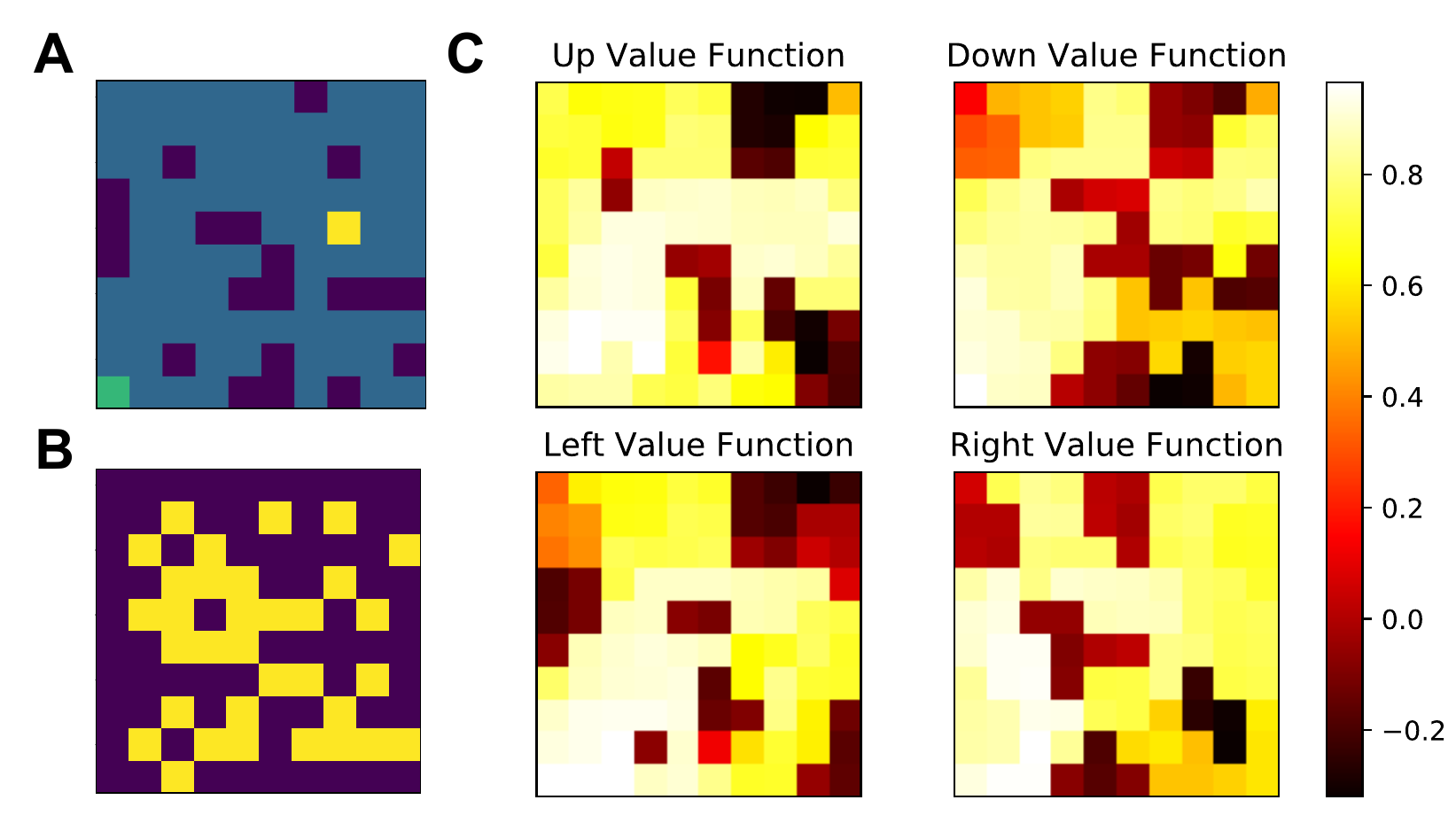}
	\caption{\textbf{A:} Randomly generated grid world, dark blue cells represent negative rewards (-1), the green cell represents the goal (+1) and the yellow cell represents the agents starting position. \textbf{B:} Image showing the locations encoded by the SOM component of CTDL (yellow cells) at the end of learning in A. \textbf{C:} CTDLs value function at the end of learning in A, the value is calculated as the weighted average of the predictions from the SOM and DNN. Each state has four possible values, corresponding to each of the four possible actions (up, down, left and right). }
	\label{fig:F2_SOMAndValue}
\end{figure}

Figure \ref{fig:F1_CTDLAndDQN} demonstrates the results of the two approaches on a random selection of grid worlds.  CTDL outperforms the DQN in terms of cumulative reward and the cumulative number of 'ideal' episodes. An ideal episode is classified as an episode where the agent avoids all negative rewards and reaches the positive reward. These findings suggest that the inclusion of a second 'hippocampal' system, which explicitly contributes to the calculation of Q values, is beneficial in our simple grid world task. This gain in performance is achieved at a much lower cost in terms of memory resources. Figure \ref{fig:F2_SOMAndValue} shows an example maze along with the weights of each unit in the SOM and the location they represent in the maze at the end of learning.

\begin{figure}
	\centering
	\includegraphics[height=12cm, width=12cm]{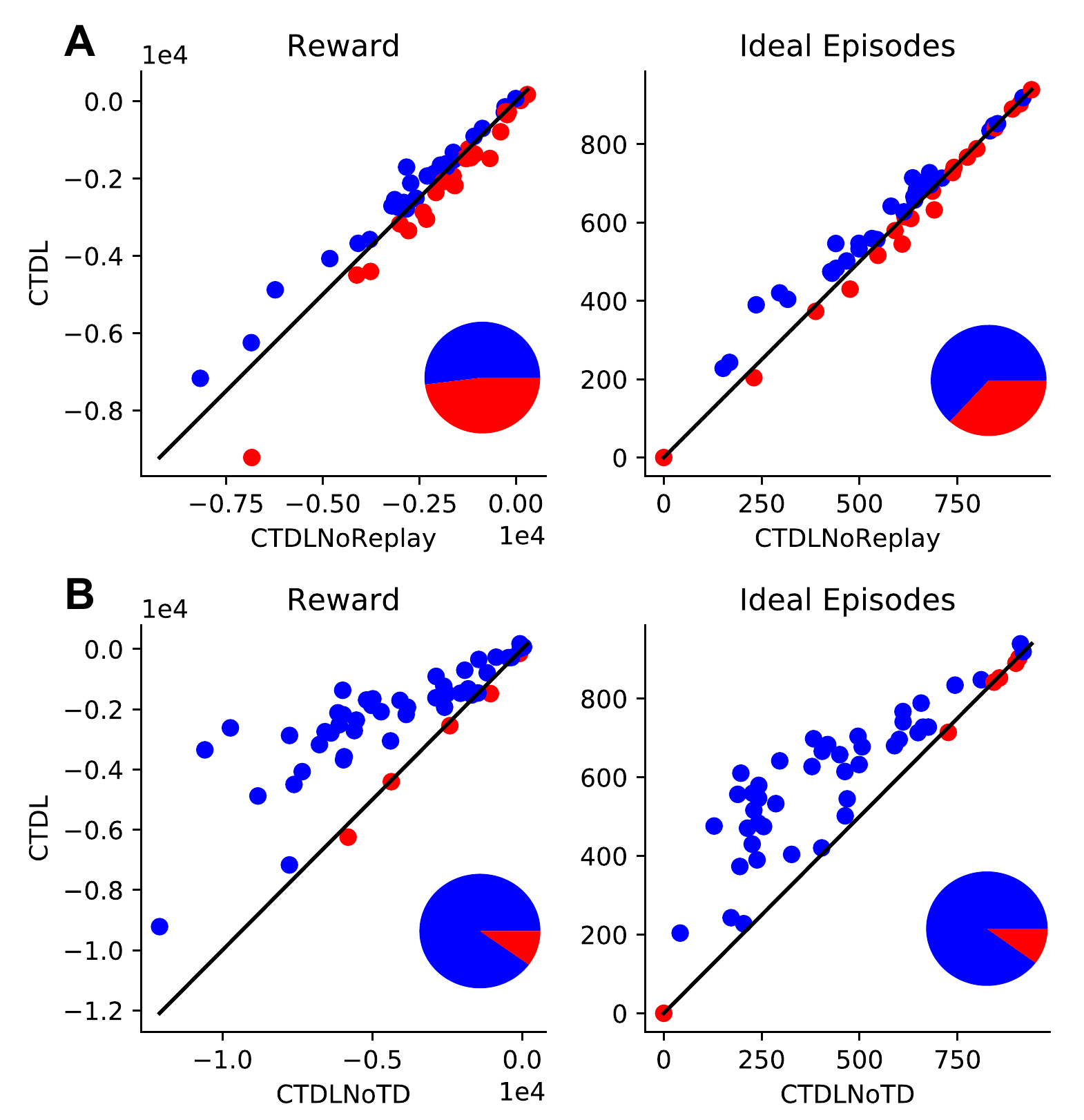}
	\caption{\textbf{A:} Scatter plots comparing the performance of CTDL and CTDL without replay on 50 different randomly generated grid worlds. Both CTDL and CTDL without replay were run 30 times on each maze. Blue points indicate grid worlds where CTDL out-performed CTDL without replay and red points indicate grid worlds where CTDL without replay out-performed CTDL. The pie chart to the lower right indicate the proportions of blue and red points. \textbf{B:} Scatter plots comparing the performance of CTDL and CTDL without TD learning in 50 different procedurally generated grid worlds. Both CTDL and CTDL without TD learning were run 30 times on each maze. Blue points indicate grid worlds where CTDL out-performed CTDL without TD learning and red points indicate grid worlds where CTDL without TD learning out-performed CTDL. The pie charts to the lower right indicate the proportions of blue and red points.}
	\label{fig:F3_ReplayAndTD}
\end{figure}

To improve our understanding of the mechanisms underlying CTDLs performance we isolated the contribution of the SOM to the calculation of the Q values from the replaying of the contents of the SOM to the DNN. Figure \ref{fig:F3_ReplayAndTD}A shows the performance of CTDL both with and without replay. CTDLs performance was only marginally reduced by the removal of replay suggesting that the improvements over the DQN are due to the contribution of the SOM to the calculation of Q values. A key component of CTDL is the updating of the SOM using the TD error from the DNN. To investigate the importance of this interaction, we compared CTDL to a version of CTDL that did not update the SOM based on the TD error from the DNN. This was achieved by setting the learning rate of the SOM to $0$ so that the weights $\beta_u$ were not updated during learning. Figure \ref{fig:F3_ReplayAndTD}B shows the results of this comparison. Removal of the interaction between the DNN and the SOM via the TD signal had a significant impact on the performance of CTDL, suggesting that it is a critical component of the model.

\begin{figure}
	\centering
	\includegraphics[height=14cm, width=11cm]{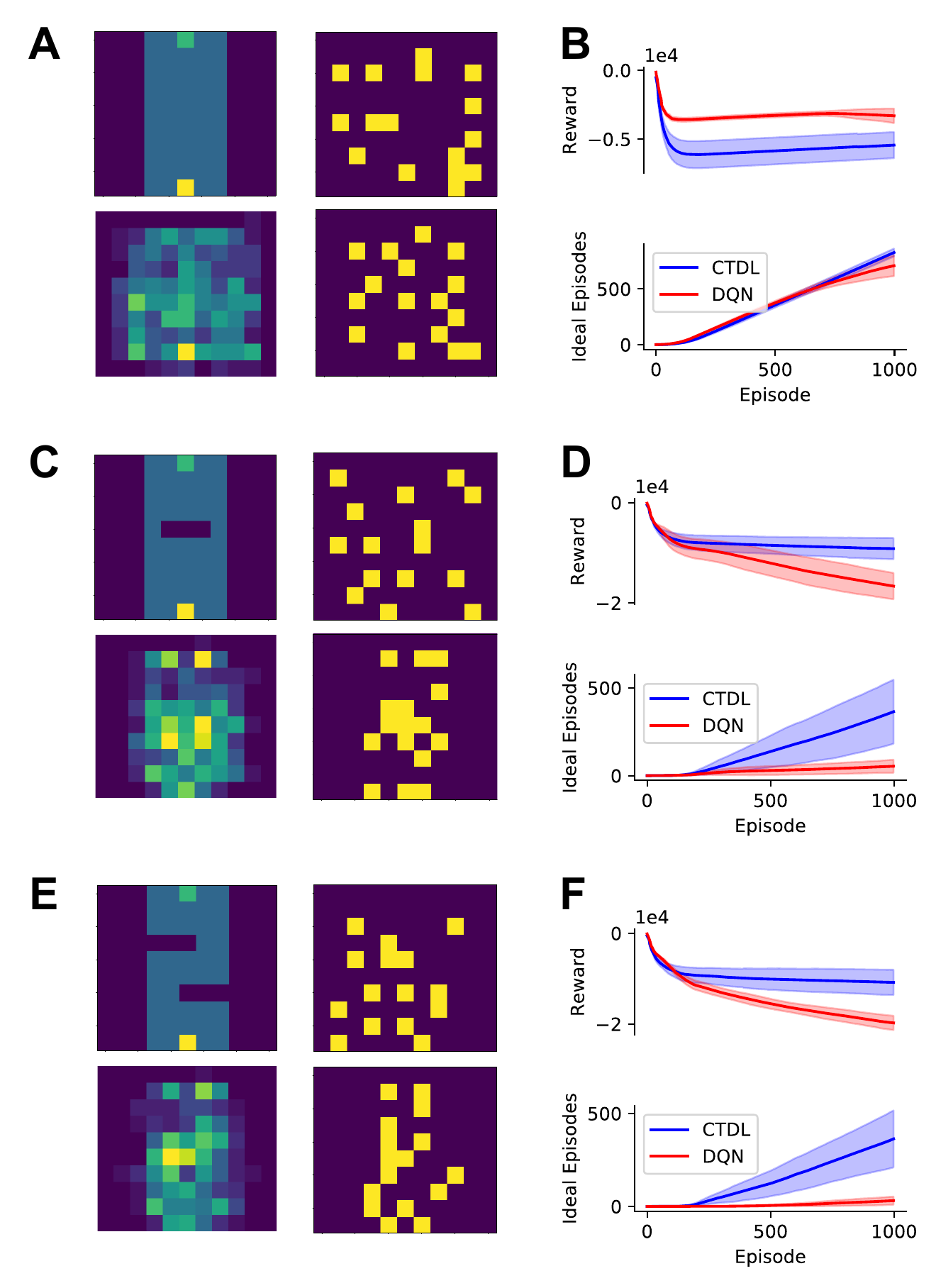}
	\caption{\textbf{A:} \textit{Top-Left:} Grid world where the agent only has to travel upwards to reach the goal. Dark blue cells represent negative rewards (-1), the green cell represents the goal (+1) and the yellow cell represents the agents starting position. \textit{Bottom-Left:} Locations encoded by the SOM component of CTDL at the end of learning, results are averaged over 30 runs. \textit{Top-Right:} Locations encoded by the SOM component of CTDL at the start of learning for a single run. \textit{Bottom-Right:} Locations encoded by the SOM component of CTDL at the end of learning for a single run. \textbf{B:} The performance of CTDL and DQN on the grid world from A in terms of cumulative reward and 'ideal' episodes. The solid line represents the mean and the shaded region represents the standard deviation. \textbf{C:} Same as A but an obstacle is introduced, in the form of negative rewards, that the agent must circumnavigate. \textbf{D:} The performance of CTDL and DQN on the grid world from C. \textbf{E:} Same as C but with two obstacles for the agent to circumnavigate. \textbf{F:} The performance of CTDL and DQN on the grid world from E.}
	\label{fig:F4_Obstacles}
\end{figure}

One interpretation of these results is that the SOM is able to store and use experiences that violate generalizations made by the DNN and that this confers a significant advantage during learning. To test this hypothesis we ran CTDL and DQN on three new mazes (Figure \ref{fig:F4_Obstacles}). The first maze had no negative rewards between the start and goal locations and the agent simply had to travel directly upwards. We predict that such a maze should favour the DQN because it can rely upon the generalization that an increase in $y$ corresponds to an increase in expected return. The second and third mazes introduced negative rewards that violate this generalization. For these mazes we predict that CTDL should perform better because it can store states that violate the generalization in its SOM and when these states are re-visited CTDL can consult the Q values predicted by the SOM. Figure \ref{fig:F4_Obstacles} shows the results of CTDL and DQN on these three mazes. To help visualise the locations encoded by the SOM we reduced the SOM size to 16 units. The results provide support for our predictions, with the DQN out-performing CTDL in the first maze but not in the second and third mazes. Interestingly, over the course of learning the locations encoded by the SOM appeared to reflect regions of the maze that correspond to violations in the 'move upwards' generalization. We take these findings as evidence that the SOM is encoding states that violate generalizations made by the DNN and that this is responsible for CTDLs improved performance.  

\begin{figure}
	\centering
	\includegraphics[height=12cm, width=10.5cm]{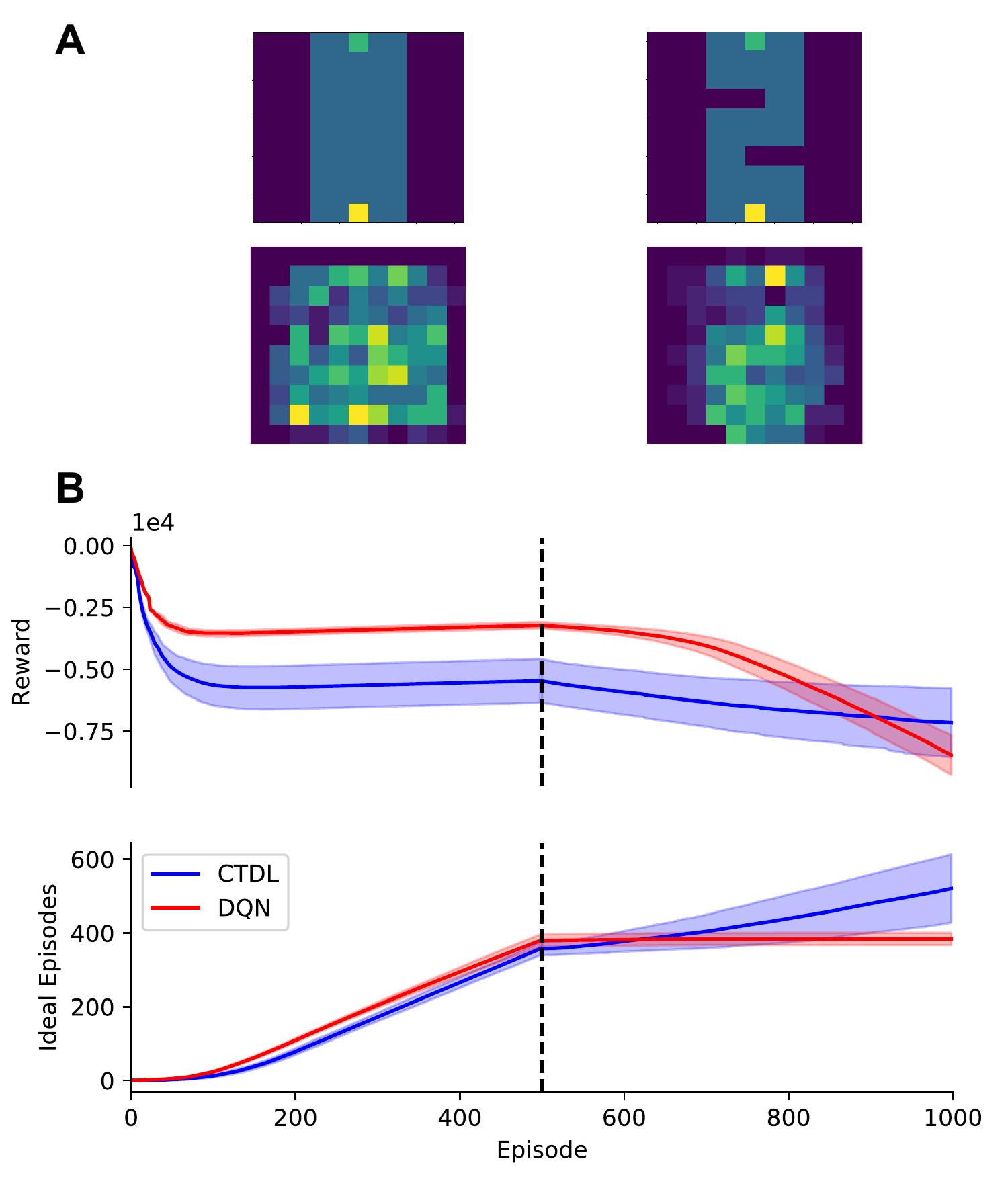}
	\caption{\textbf{A:} \textit{Top-Left:} First grid world presented to the agent for 500 episodes. \textit{Bottom-Left:} Locations encoded by the SOM component of CTDL at the end of learning in the first grid world, results are averaged over 30 runs. \textit{Top-Right:} Second grid world presented to the agent for 500 episodes immediately after the first grid world. \textit{Bottom-Right:} Locations encoded by the SOM component of CTDL at the end of learning in the second grid world, results are averaged over 30 runs. \textbf{B:} The performance of CTDL and DQN on the successive grid worlds from A. The solid line represents the mean and the shaded region represents the standard deviation. The dashed line indicates the change in grid worlds and the introduction of the obstacles.}
	\label{fig:F5_Revaluation}
\end{figure}

If the SOM does encode states that violate generalizations made by the DNN, then this should translate to improved behavioral flexibility in the face of environmental changes. For example if an obstacle appears in one of the grid worlds then this should lead to a large TD error and instruct the SOM to encode the position of the obstacle using its large learning rate. Subsequently, since the SOM keeps track of action values independently from the DNN, CTDL should be able to quickly adapt its behavior in order to avoid the obstacle. To investigate this hypothesis we ran CTDL and DQN on the grid world in Figure \ref{fig:F4_Obstacles}A immediately followed by the grid world in Figure \ref{fig:F4_Obstacles}C. Figure \ref{fig:F5_Revaluation} shows the results of these simulations. As previously described, the DQN out-performed CTDL on the first grid world in terms of cumulative reward and the number of ideal episodes. Switching to the second grid world impacted the performance of both the DQN and CTDL. However, this impact was more pronounced for the DQN, with a larger decrease in cumulative reward and a plateauing of the number of ideal episodes. This suggests that CTDL is better equipped to handle changes in the environment. As before the locations encoded by the SOM appeared to reflect states immediately preceding the obstacle. This is consistent with the notion that the TD error from the DNN allows the SOM to identify regions that violate the generalizations made by the network and subsequently improve learning.

One of the strengths of RL algorithms is that they can be applied to a wide array of tasks. If one can describe a task using a state space, an action space and a reward function then often it can be solved using RL techniques, especially if the states satisfy the Markov property. We therefore wanted to investigate whether the performance of CTDL was specific to grid worlds or whether it could be applied to another task. We chose to test CTDL on the Cart-Pole environment from OpenAI Gym. We chose the Cart-Pole environment because it is a common benchmark task in the RL literature and it involves a continuous state space, unlike the discrete state space of the grid world environments. The parameter values used for all Cart-Pole simulations were the same as in the grid world simulations with two exceptions. Firstly, the number of time steps $C$ between updates of the target network was changed to 500 in order to account for the shorter episodes experienced in the Cart-Pole task. Secondly, the size of the SOM was increased from 36 units to 225 units, which is still considerably smaller than the size of the replay buffer used by the DQN (100,000). 

An important component of CTDL is the calculation of the euclidean distance between the current state $s_t$ and the weights of each unit $\beta_u$. In the case of the Cart-Pole task this will cause the velocity values in the state representation to dominate the distance calculations because their values cover a much greater range. To account for this we maintain an online record of the largest and smallest values for each entry in the state representation and use these values to normalise each entry so that they lie in the range $[0, 1]$. This ensures that each entry in the state representation contributes equally to any euclidean distance calculations.

\begin{figure}
	\centering
	\includegraphics[height=6cm, width=6cm]{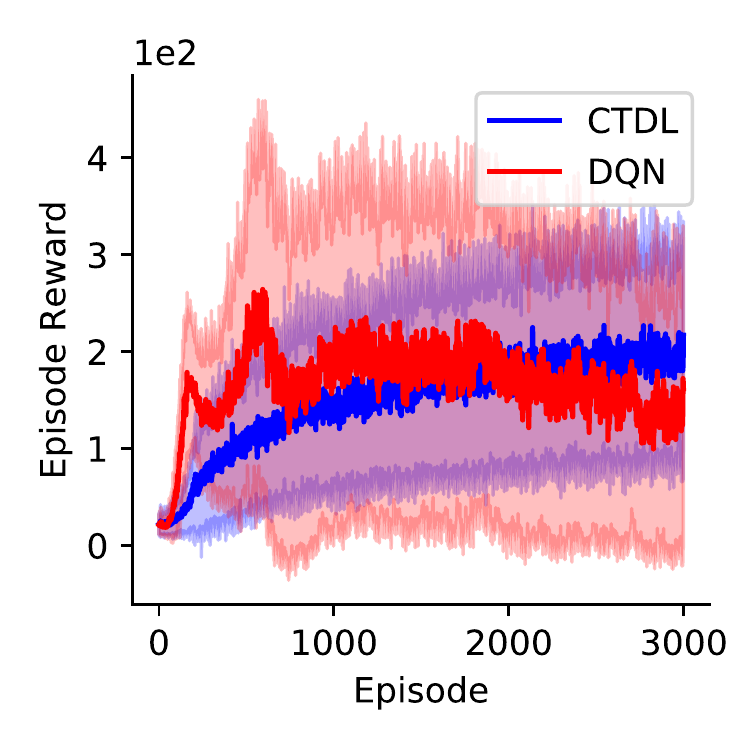}
	\caption{Episode reward achieved by CTDL and DQN on the Cart-Pole environment over the course of learning. Both CTDL and DQN were run 100 times on the Cart-Pole environment. The solid line represents the mean and the shaded region represents the standard deviation.}
	\label{fig:F6_CartPole}
\end{figure}

Figure \ref{fig:F6_CartPole} shows the results of both CTDL and DQN on the Cart-Pole task. While the DQN appeared to learn faster than CTDL, it did so with greater variance and the stability of the final solution was poor. In comparison, CTDL learnt gradually with less variance and there were no significant decreases in performance. These results demonstrate that CTDL can be applied to continuous state problems and is not restricted to discrete grid world problems. They also suggest that CTDL's use of dual learning systems may confer a stability advantage that improves the robustness of learning.

\section{Neural Underpinnings}

It is important to consider how our proposal maps onto real neural systems. As previously mentioned, the key components of CTDL are the independent contribution of a 'hippocampal' learning system to the evaluation of states and actions and the use of TD error to update representations in the 'hippocampal' learning system. With respect to the first of these components, it is well known that the striatum is a central location for updating and evaluating states and actions for decision making \citep{Schultz1992, Houk1995, Schultz1998, Setlow2003, Roesch2009}. Importantly, the striatum receives direct inputs from both cortical areas and the hippocampus \citep{Groenewegen1987, Thierry2000}. It has also been proposed that pattern-separated hippocampal representations aid the reinforcement learning process \citep{Duncan2018, Ballard2019}. These findings lend support to the idea that a 'hippocampal' learning system may provide a value prediction that complements that of a 'neocortical' learning system. In addition, with converging cortical and hippocampal inputs, the striatum needs to be able to arbitrate between them in order to calculate a state or action value \citep{Pennartz2011}. The weighting process between the SOM and DNN components performed by CTDL may represent a simplified example of such an arbitration. Interestingly, CTDL predicts that the striatum should apply a greater weighting to hippocampal information when the current state closely matches one stored in episodic memory. 

The second component of CTDL relies on the use of TD error to update both hippocampal value estimates and memory representations. A wealth of evidence currently suggests that the primary neural correlates of TD error are phasic dopamine neurons in the midbrain \citep{Schultz1997, Schultz2016}. One of the major projection sites of these neurons is the striatum and evidence of TD error has subsequently been found in the striatum \citep{Doherty2003, Mcclure2003, Bray2007}. It therefore seems plausible that TD error can be used to update value estimates situated at hippocampal-striatal synapses that are independent of neocortical value estimates. 

Nevertheless, the question remains whether TD error could modulate pattern-separated state representations in the hippocampus, as in CTDL. It has been widely reported that midbrain dopamine neurons project directly to the hippocampus and can influence synaptic plasticity in the hippocampus via Long-Term Potentiation (LTP) \citep{Lisman2005, Lemon2006, Rosen2015}. It is therefore believed that midbrain dopamine neurons can mediate the formation of episodic memories in order to guide memories towards experiences that are relevant for behavior \citep{Shohamy2010}. This biasing of episodic memory towards reward-related experiences can take many forms. For example, reward cues appear to engage midbrain dopamine neurons, which then enhances episodic memory for those cues \citep{Wittmann2005}. Similarly, motivation to obtain future rewards also promotes the firing of midbrain dopamine neurons and subsequent episodic memory of items, even in the absence of reward during learning \citep{Adcock2006}. 

While these results lend support to the hypothesis that midbrain dopamine neurons can bias the formation of episodic memories in the hippocampus, they do not provide evidence that reward prediction errors, such as TD error, have an effect. Dopaminergic midbrain neurons are thought to encode many aspects of reward-related information such as reward outcome, expected reward, novelty and incentive salience \citep{Shohamy2010}. Part of the ability of dopaminergic neurons to encode these different forms of information may lie in the differences between tonic and phasic dopamine responses. It is likely that many of the aforementioned effects on episodic memory are due to the tonic responses of dopamine neurons that encode reward-related information other than prediction errors \citep{Shohamy2010}.

CTDL specifically predicts that reward prediction errors, as encoded by phasic midbrain dopamine neurons, should promote the formation of episodic memories in the hippocampus. Empirical support for such a prediction is beginning to emerge. In particular, a recent study by \citet{Rouhani2018} demonstrated that unsigned reward prediction errors enhance episodic memory for trial-unique images. When the reward outcome differed by a large amount from the participant's subjective expected value of an image, the participant was better at recognizing that image in a subsequent surprise recognition test. This effect was consistent even when controlling for reward outcome and subjective value estimates. The effect was also independent of sign (i.e. both large positive and negative reward prediction errors improved episodic memory for images that lead to the reward prediction error), which is consistent with CTDL as the algorithm uses the absolute value of the TD error to update the SOM. Interestingly, \citet{Rouhani2018} also found that when participants were presented with the same images again, they tended to choose the one that previously had the larger reward outcome. This suggests that they also encoded the rewards associated with the images. From the perspective of CTDL this could be seen as encoding the value of the episodic memory independently from the 'neocortical' learning system. Indeed, the result of the parameter sweep assigned a large value to the learning rate ($\rho = 0.9$, Table \ref{table:HyperParameterTable}) for the Q values of the SOM, perhaps reflecting a direct episodic encoding of the reward outcome rather than a running average. 

Further evidence for the promotion of episodic memory via reward prediction errors comes from a study by \citet{Jang2018}. In this study, participants had to decide whether to play or pass on a risky gamble. To make the decision participants were provided with information about the potential payout and an image from one of two categories, which they could use to incrementally learn the reward probability of that category. If the participants chose to play the risky gamble then the subsequent feedback was active, otherwise it was passive. Importantly, the authors showed that episodic memory for images was improved when reward prediction errors were large at the time of image presentation. This effect was only apparent for active as opposed to passive feedback, suggesting that it was dependent on decision-making. In addition, the effect was consistent regardless of whether the image recognition task was performed immediately after the decision-making task or 24 hours after. This suggests that the effect of reward prediction errors on episodic memory and subsequently decision-making are potentially fast acting and do not require consolidation mechanisms. Taken together these findings provide additional support for the modulation of episodic memory formation via reward prediction errors.

Reward prediction errors have also been proposed to have a role in the updating of long-term memories. The theory of memory reconsolidation posits that long-term memories which have been destabilised into a malleable form can be updated with new information to aid integration and avoid interference \citep{Sara2000}. This process is thought to be hippocampus-dependent \citep{Debiec2002, Lee2004} and rely upon reward prediction errors from midbrain dopamine neurons to signal the need for integration of new information \citep{Exton-mcguinness2015}. Evidence for this comes from the fact that reconsolidation appears to decrease once a behavior becomes well-learnt, supposedly because reward prediction errors have decreased. This provides an example of how reward prediction errors, such as TD error, may be able to modify existing memories via the hippocampus; a process that is critical to CTDL.

While the aforementioned studies demonstrate that reward prediction errors, such as TD error, can modulate episodic memory, it is worth noting that these findings are not unanimous. Most notably an fMRI study by \citet{Wimmer2014} found a negative correlation between striatal reward prediction errors and performance on an episodic memory task. The authors suggested that the formation of episodic memory interferes with classical reinforcement learning by disrupting reward prediction errors. Such a finding argues against the prediction of CTDL that TD error should promote the formation of episodic memories, however criticisms of the study have been highlighted. In particular, the trial-unique images were unrelated to the actual reward learning task and so the reward prediction errors may not have been elicited by the images themselves. In comparison, the studies by \citet{Rouhani2018} and \citet{Jang2018} both required participants to use the images to perform the decision-making task. In addition, it would have been interesting to see the results of the correlation analysis using absolute or unsigned reward prediction errors. Both CTDL and \citet{Rouhani2018} predict no linear relationship between signed reward prediction errors and episodic memory performance. Nevertheless, such confounding findings highlight the need for further empirical work to elucidate the role of reward prediction errors in the formation of episodic memories.

In reality, it is likely that a combination of reward-related information, encoded by midbrain dopaminergic neurons, is responsible for the dynamic and selective episodic memory present in humans. For example a study by \citet{Mason2017} explored the effect of different reward-related effects on episodic memory. In the study participants had to learn a collection of words in exchange for monetary rewards, thereby creating a motivated learning scenario. The authors found that reward outcome i.e. a combination of expected value and reward prediction error, was the best predictor of episodic memory. This demonstrates that reward prediction error on its own may not provide a holistic account of episodic memory formation. Taking this into account, we believe that CTDL provides a useful initial framework for exploring the effect of other reward-related signals on the formation of episodic memories and subsequently goal-directed behavior.

In addition to the algorithmic components, the behavior demonstrated by CTDL has several interesting parallels with biological findings. Firstly, the 'hippocampal' learning system (i.e. the SOM) of CTDL appears to encode violations of the generalizations made by the DNN. In the case of grid worlds this corresponded to regions close to obstacles. This finding has an interesting parallel with imaging work in rodents demonstrating that CA3 neurons appear to encode decision points in T-mazes that are different from the rodents current position \citep{Johnson2007}. Such decision points could be viewed as obstacles or important deviations from the animals general direction. Their encoding by the hippocampus is therefore consistent with being encoded by the SOM component of CTDL. In the future, application of CTDL to other reinforcement learning tasks may make testable predictions about the regions of the state space that should be encoded by the hippocampus. Another key behavior demonstrated by CTDL was increased flexibility when presented with a change in the environment i.e. the introduction of an obstacle. This was due to the ability of the SOM to quickly encode the states close to the obstacle. This suggests that the hippocampus may be important for adapting to changes in the environment and is consistent with recent studies that have implicated the hippocampus in reversal learning \citep{Dong2013, Vila-Ballo2017}.

Aside from 'what' should be encoded by the hippocampus, CTDL is also consistent with biological theories regarding the 'duration' of encoding. It has been proposed that memories stored in the hippocampus are consolidated to the neocortex over time via mechanisms such as replay \citep{Olafsdottir2018}. Interestingly, this process should naturally occur in CTDL; as the neural network improves its ability to evaluate the optimal value function the TD errors should reduce in magnitude and free up the SOM to represent other episodic memories. If part of the environment changes then a new episodic memory will form based on the TD error and it will remain in episodic memory until the 'neocortical' learning system has learnt to incorporate it. This process suggests that the transfer of information from the hippocampus to the neocortex is very much related to the 'need' for an episodic memory as encoded by TD errors.

\section{Discussion}

According to CLS theory, the brain relies on two main learning systems to achieve complex behavior; a 'neocortical' system that relies on the slow learning of distributed representations and a 'hippocampal' system that relies on the fast learning of pattern-separated representations. Both of these systems project to the striatum, which is believed to be a key structure in the evaluation of states and actions for RL \citep{Schultz1992, Houk1995, Schultz1998, Setlow2003, Roesch2009}. Current deep RL approaches have made great advances in modelling complex behavior, with DNNs sharing several similarities with a 'neocortical' learning system. However these approaches tend to suffer from poor data efficiency and general inflexibility \citep{Lake2017}. The purpose of the present study was to explore how a 'neocortical' and 'hippocampal' learning system could interact with the brain's RL machinery and whether CLS theory could alleviate some of the criticisms of deep RL.

Our novel approach, termed CTDL, used a DNN as a 'neocortical' learning system and a SOM as a 'hippocampal' learning system. Importantly the DNN used a small learning rate and distributed representations while the SOM used a larger learning rate and pattern-separated representations. Our approach is novel in that the SOM contributes to action value computation by storing action values independently from the DNN and uses the TD error produced by the DNN to update its state representations. More specifically, the TD error produced by the DNN is used to dynamically set the learning rate and standard deviation of the neighbourhood function of the SOM in an online manner. This allows the SOM to store memories of states that the DNN is poor at predicting the value of and use them for decision-making and learning. Importantly the size of the SOM is smaller than the state space encountered by the agent and so it requires less memory resources than the purely tabular case.

We compared the performance of CTDL to a standard DQN on a random set of 2D grid worlds. CTDL out-performed the DQN on the majority of grid worlds, suggesting that the inclusion of a 'hippocampal' learning system is beneficial and confirming the predictions of CLS theory. Removal of replay between the SOM and DNN appeared to have marginal impact upon the performance of CTDL suggesting that the SOMs contribution to the calculation of action values is the predominant benefit of CTDL. Future work should explore how information from the SOM may be replayed to the DNN in a more principled fashion (e.g. \citet{Mattar2018}) instead of random sampling. We proposed that the SOM was able to contribute to the calculation of the action values in a targeted manner by using the TD error of the DNN to encode states that the DNN was poor at evaluating. We provided evidence of this by demonstrating that the removal of the TD signal between the DNN and SOM had a negative impact upon the performance of CTDL. 

Our interpretation of these results is that, particularly early on in learning, the DNN is able to represent generalizations of the state space while the SOM is able to represent violations of these generalizations. In combination these two systems can then be used to formulate policies in both a general and specialised manner. We tested this hypothesis by presenting CTDL and DQN with a grid world consisting of a general rule and two other grid worlds consisting of violations of this rule. As our interpretation predicted, CTDL out-performed the DQN when violations of the general rule were present, presumably because the SOM was able to store states that were useful for circumnavigating these violations. This hypothesis was further supported by a simulation that ran both CTDL and DQN on sequential gird worlds. CTDL appeared to be better equipped to deal with the change in environment compared to the DQN. In addition, the SOM component of CTDL encoded states close to the change in the environment, providing further evidence of its ability to represent violations of predictions.

To investigate the generality of CTDL we also applied it to the Cart-Pole problem, which is characterised by a continuous state space. We found that in comparison to the DQN, the learning of CTDL was more gradual but also more robust. This is perhaps a surprising result given that the DQN has a perfect memory of the last 100,000 state transitions whereas CTDL has no such memory. Indeed, one would expect the SOM component of CTDL to have less of an effect in continuous state spaces because generalization from function approximation becomes more important and the probability of re-visiting the same states decreases. With this being said, \citet{Blundell2016} demonstrated that even when the probability of re-visiting the same state is low, episodic information can still be useful for improving learning. Generalization of episodic information in CTDL is likely controlled by the temperature parameter $\tau_{\eta}$ that scales the euclidean distance between the states and the weights of the SOM units. 

Further work will need to investigate whether the increased robustness of CTDL in continuous state spaces is a general property that applies to a variety of RL tasks. In particular, it would be of interest to run CTDL on maze problems such as ViZDoom \citep{Kempka2016}, which are rich in visual information.  Indeed, deep RL approaches using convolutional neural networks are at the forefront of RL research and these could be easily incorporated into the CTDL approach. In the case of ViZDoom, each state is represented by a high-dimensional image and so the generalization capabilities of a DNN are crucial. From a biological perspective, it is worth noting that the hippocampus operates on cortical inputs that provide latent representations for episodic memory. This is captured in 'model-free episodic control', which relies on an embedding function to construct the state representation for episodic memory \citep{Blundell2016, Pritzel2017}. An embedding function therefore represents a biologically plausible method of scaling CTDL up to complex visual problems such as VizDoom. The embedding function could be pre-trained in an unsupervised manner or sampled from the DNN component of CTDL. We leave this interesting avenue of research to future work.

One consistent feature of the tasks presented in the present study is their low degree of stochasticity. As with discrete state spaces, low stochasticity means that events re-occur with high probability and the episodic component of CTDL can exploit this. It is likely that in more stochastic environments the benefits of CTDL will be reduced as the DNN is required to generalise over several outcomes. It is therefore an open question how well CTDL will perform on tasks that have a high degree of stochasticity, which are also supposedly harder for biological agents.

One of the primary goals of the present study was to explore a biologically plausible implementation of CLS theory within an RL framework. The central components of CTDL are a 'neocortical' and 'hippocampal' learning system that both contribute to the calculation of action values and the use of TD error to update the 'hippocampal' learning system. We believe both of these components may be represented in the brain through interactions between the neocortex, hippocampus, striatum and midbrain dopamine neurons. In particular, midbrain dopamine neurons encode TD error and project to both the striatum and hippocampus. These projections may be responsible for the modulation of action values via neocortical-striatal and hippocampal-striatal synapses and the modulation of episodic memory content via intra-hippocampal synapses.

This biological interpretation raises several important empirical predictions. Most notably, CTDL predicts that absolute unsigned reward prediction errors should promote the formation of episodic memories. Empirical evidence for such a phenomenon is still tentative and so we highlight this as a critical area for future empirical investigation. Findings from neuroscience research suggest that other reward-related signals from midbrain dopaminergic neurons may also influence the content of episodic memory. The addition of these other reward-related signals represent potential extensions to the CTDL framework and may provide additional learning advantages to an agent.

In the present study we have only demonstrated the benefits of a 'hippocampal' learning system from a purely model-free perspective. A growing body of research however, is implicating the hippocampus in what has historically been considered model-based behavior. For example, it has been proposed that the hippocampus encodes a predictive representation of future state occupancies given the current state of the agent \citep{Stachenfeld2017}. This predictive representation has been termed the Successor Representation (SR), and has been shown to imbue agents with model-based behavior using model-free RL mechanisms in a range of re-evaluation tasks \citep{Russek2017}. Therefore the inclusion of a 'hippocampal' learning system may have additional benefits to those demonstrated by CTDL.

\section{Conclusions}

Taken together, we believe that our work highlights CTDL as a promising avenue for achieving complex, human-like behavior and exploring RL within the brain. Having a 'neocortical' and 'hippocampal' learning system operating in parallel conferred a learning advantage over a single 'neocortical' system. This advantage was attributable to two main properties of CTDL. Firstly, both the 'neocortical' and 'hippocampal' system contributed to the calculation of action values for decision-making, with the arbitration between the two dependent on the memory content of the 'hippocampal' system. Secondly, the contents of the 'hippocampal' system were dynamically updated using the TD error from the 'neocortical' system. This allowed the 'hippocampal' system to target regions of the state space that the 'neocortical' system was poor at evaluating or that violated generalizations made by the 'neocortical' system. 

These key properties of CTDL represent promising avenues for future research both computationally and empirically. From a computational perspective, it will be interesting to explore how embedding functions can be utilised to reflect the fact that the hippocampus receives latent representations from cortical areas as input. This may be a key component for scaling up CTDL to complex problems with high dimensional state spaces. With respect to future empirical work, CTDL specifically predicts that unsigned reward prediction errors should promote the formation of episodic memories in the hippocampus. Conflicting evidence currently exists for this prediction and so we highlight this as a key area for further investigation and clarification.

\section*{Acknowledgments}

This work was funded by the UK Biotechnology and Biological Sciences Research Council (BBSRC). We thank NVIDIA for a hardware grant that provided the Graphics Processing Unit (GPU) used to run the simulations.

\section*{References}
\bibliographystyle{agsm}
\bibliography{CTDL}

\end{document}